\documentclass[%
 amsmath,amssymb,
prf,
notitlepage
]{revtex4-1}
\usepackage{layout}
\usepackage{graphicx,color}
\usepackage{amsmath,amsfonts,amssymb}
\usepackage{bm}
\usepackage[symbol]{footmisc}
\usepackage{epstopdf}
\usepackage[section]{placeins}
\usepackage[font={footnotesize}]{caption}
\usepackage{ulem}

\newcommand{\bcdot}{\bm\cdot}
\newcommand{\bnabla}{\bm\nabla}
\newcommand{\bPi}{\bm\Pi}
\newcommand{\bmu}{\bm u}
\newcommand{\bmQ}{\mathrm{\mathbf{Q}}}
\newcommand{\bmC}{\mathrm{\mathbf{C}}}
\newcommand{\bmS}{\mathrm{\mathbf{S}}}
\newcommand{\bmH}{\mathrm{\mathbf{H}}}
\newcommand{\bmD}{\mathrm{\mathbf{D}}}
\newcommand{\bmI}{\mathrm{\mathbf{I}}}
\newcommand{\trace}[1]{\operatorname{Tr}\left[#1\right]}
\newcommand{\Em}[1]{{\color{black}#1}}

\newcommand{\nn}{\nonumber}

\begin{document}
\title{Active matter in a viscoelastic environment}
\author{Emmanuel L. C. VI M. Plan}
\affiliation{Institute of Theoretical and Applied Research, Duy Tan University, Hanoi 100000, Viet Nam}

\author{Julia M. Yeomans}
\affiliation{The Rudolf Peierls Centre for Theoretical Physics, Department of Physics, University of Oxford,\\ Clarendon Laboratory, Parks Road,  Oxford OX1 3PU, United Kingdom}

\author{Amin Doostmohammadi}
\affiliation{Niels Bohr Institute, Blegdamsvej 17, 2100, Copenhagen, Denmark}

\begin{abstract}
Active matter systems such as eukaryotic cells and bacteria continuously transform chemical energy to motion. Hence living systems exert active stresses on the complex environments in which they reside.
One recurring aspect of this complexity is the viscoelasticity of the medium surrounding living systems: bacteria secrete their own viscoelastic extracellular matrix, and cells constantly deform, proliferate, and self-propel within viscoelastic networks of collagen. 
It is therefore imperative to understand how active matter modifies, and gets modified by, viscoelastic fluids. 
Here, we present a two-phase model of active nematic matter that dynamically interacts with a passive viscoelastic polymeric phase \Em{and perform numerical simulations in two dimensions to illustrate its applicability}. 
Motivated by recent experiments we first study the suppression of cell division by a viscoelastic medium surrounding the cell. 
We further show that the self-propulsion of a model keratocyte cell is modified by the polymer relaxation of the surrounding viscoelastic fluid in a non-uniform manner and find that increasing polymer viscosity effectively suppresses the cell motility. 
Lastly, we explore the hampering impact of the viscoelastic medium on the generic hydrodynamic instabilities of active nematics by simulating the dynamics of an active stripe within a polymeric fluid. 
The model presented here can provide a framework for investigating more complex dynamics such as the interaction of multicellular growing systems with viscoelastic environments.
\end{abstract}

\maketitle

\section{Introduction}
Active matter describes a class of living systems such as cellular tissues, bacterial colonies, and subcellular filaments that are continuously put in motion by the activity of their building blocks: each individual cell in the tissue or \Em{each} bacterium works as a machine, actively converting the chemical energy of the environment to mechanical work~\cite{MJRLPRA13,Bechinger16,DIYS18}.
The majority of active matter research has focused on 
how these entities navigate through and exhibit collective dynamics in a viscous fluid.
Another important factor, however, is the elasticity of the fluid because it can give rise to a completely different dynamics \cite{EL15,EWG15} that, in extreme cases, may dictate the preservation or the annihilation of a population.
For instance bacterial colonies on a surface, \textit{e.g. Escherichia coli}, aggregate into a biofilm by secreting exopolysaccharides. 
This biofilm provides the colonies with additional tolerance from external physical forces or antibiotics \cite{FWSSRK16}.
Animal and human cells also encounter viscoelasticity, such as an extracellular matrix or a heterogenous mixture of stiff or elastic cells \cite{PKY14,PJJ15}.
Recent experiments highlight the role of mechanical stresses on the self-induced growth of an \textit{E. coli} biofilm in the renal system \cite{CKCG18} and in the failure of pre-mitotic elongation of cancerous cells in stiff hydrogels \cite{WWCLTMEO17,NC18}. 
It is therefore of substantial biophysical importance to understand the interaction of active matter with a surrounding viscoelastic environment.

Distinct and unpredictable behaviours are observed when the environment of an active particle is viscoelastic.
At the individual level, both decreased \cite{SA11} and enhanced \cite{BT79,PGGA15} swimming speeds are displayed by microscopic organisms in complex 
fluids\Em{, and there is a widespread agreement that the swimming speed is strongly dependent on both particle geometry and fluid rheology \cite{SLP13}, with evidence from
both theory and simulations \cite{ZLB12,SLP13,EL15,DNHE17,ZY18}.
}
Numerical studies in a Poiseuille flow show that swimmers in a non-Newtonian fluid migrate upstream faster or slower than their Newtonian counterparts depending on the shear-thickening/thinning properties of the fluid \cite{MSYD16}.
Particle trajectories also differ in polymer solutions as illustrated by the suppression of tumbling reorientations of \textit{E. coli} bacteria \cite{PGGA15} and in the preferential circular motion of synthetic light-activated swimmers \cite{NBG18}.
Moreover, when subjected to an external shear, spherical pusher (puller)-type swimmers in a complex fluid are predicted to reorient their swimming directions towards the vorticity axis for very-weakly (strongly)-elastic-fluids \cite{DD17}.

\begin{figure}[t]
\centering
\includegraphics[width=0.5\textwidth]{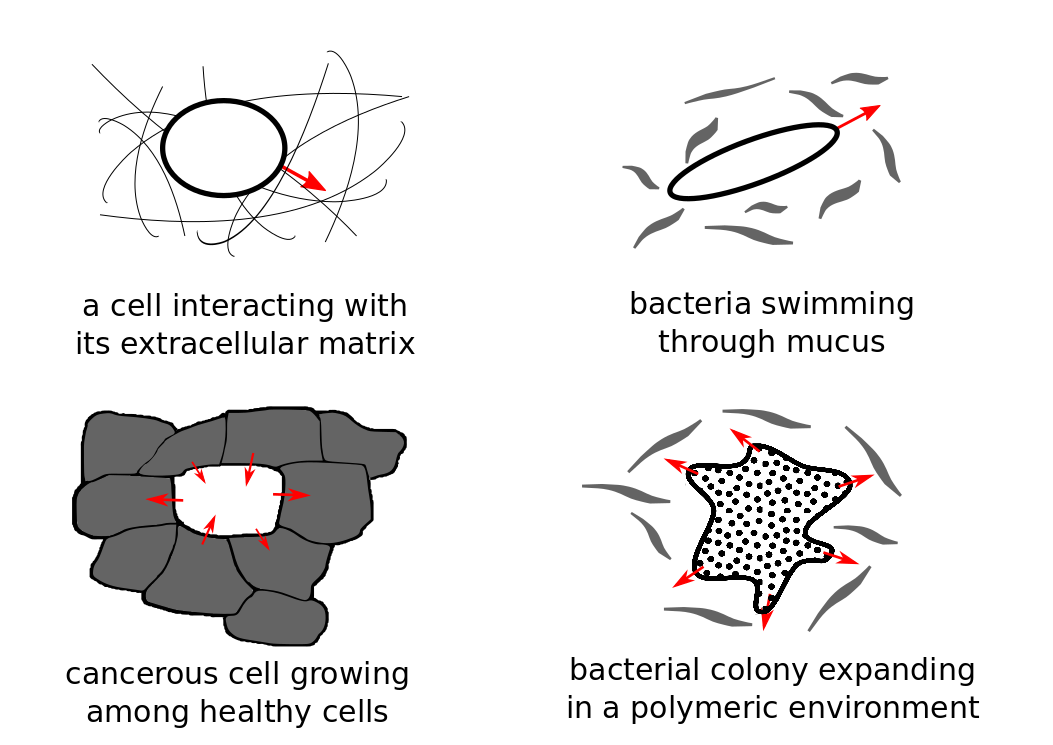}
\caption{A schematic of various interpretations of the two-phase active-viscoelastic model.}
\label{fig:schematic}
\end{figure}

Little is known about the impact of viscoelasticity on the collective behaviour of active matter. 
Synchronization and flocking have been experimentally observed in bovine sperms in viscoelastic fluids \cite{WCGR09,TLHFAMS17} and these effects were captured in simplified settings. Synchronization was replicated by idealizing sperm tails in the form of two Taylor sheets \cite{EL15}. Flocking was meanwhile achieved numerically in an assembly of discrete self-propelled extensile rods that push the fluid along their elongation axis. The rods aggregate in non-Newtonian fluids more strongly than in Newtonian fluids,
but form smaller collective structures because the resulting viscoelastic flow suppresses velocity fluctuations \cite{BU14,LA16}.
Moreover, a continuum model of active rods with viscoelastic properties has been shown to generate transient active turbulence and lead to a drag reduction in active flows~\cite{HMBMRFC15,HCF16}.
Despite these recent advances in modeling viscoelastic active matter, the physics of two-phase interactions between an active entity and a viscoelastic surrounding - for example a deformable cell propelling through the viscoelastic extracellular matrix or a growing cell colony invading a viscoelastic environment (see Fig.~\ref{fig:schematic})- is yet to be explored.

In this article we propose a generic two-phase model of active matter in a viscoelastic fluid with both activity and elasticity captured at the continuum level. 
The active matter is described by a concentration phase field and an orientational order parameter represents elongated active particles.
In a viscous environment this model captures the interfacial instabilities of cellular monolayers and the finger-like protrusions at the periphery of growing cell colonies \cite{DTSLLY15,DTY16}.
Here we surround the active material with a viscoelastic environment, which is modelled as a polymeric phase external to the active liquid crystalline phase. 
The simple constitutive model that describes the polymer rheology has been shown to quantitatively reproduce the stress measurements in elastic gels \cite{WWCLTMEO17}.
  
In section 2 we describe the equations of motion of the model and the details of the numerical simulations. 
We then use the approach to predict the failure of cell division in stiff hydrogels and the retarded motion of fish keratocyte cells in \Em{mucus} in section 3. 
We next elaborate how an elastic medium suppresses the dynamics of active matter. 
Lastly, we describe how the model might be further developed.

\section{Model}
A convenient method to describe active matter surrounded by a viscoelastic fluid is to use a phase-field model where the active matter corresponds to one phase while the viscoelastic fluid corresponds to the other. 
The shape and active dynamics of the biological material can be mimicked by coupling the active matter phase to an order parameter that obeys some phenomenological equations (\textit{e.g.} to  polarity or internal chemical concentration). 
The viscoelastic phase is captured by a polymer model via a polymer conformation tensor that has intrinsic viscous and elastic properties.
Both phases are subjected to a flow, which is governed by the Navier-Stokes equations.
Phase-field models have been used to model active droplets \cite{TMC12,GD14}, crawling cells \cite{TMC12,ZA13,A16}, and cellular or bacterial colonies \cite{DTSLLY15,DTY16,MYD19} in an isotropic fluid, but to the best of our knowledge, have never been used to examine the viscoelastic effects of a surrounding fluid using established polymer models at the continuum level. 
We note, however, that a cell has previously been described in Ref.~\cite{ZA13} to sit \textit{on} an elastic substrate. 
A framework deriving the velocity field and elastic contributions from first principles has also been proposed for active polar matter in a viscoelastic medium using a phase-field model \cite{PSB16}.

A two-phase nematohydrodynamics model has been used to study the physics of cell motility~\cite{GD14}, cell division~\cite{GD14,Marchetti17}, and growing colonies of elongated bacteria \cite{DTSLLY15,DTY16}. 
Within this framework the active phase is modelled as a nematic liquid-crystal that is in contact with an isotropic fluid. 
The nematic symmetry is motivated by compounding evidence in various cellular and subcellular systems such as microtubule/kinesin motor mixtures, acto-myosin complexes, rod-shaped bacterial colonies, spindle-shaped fibroblasts, and neural progenitor stem cells, which all show defining features of nematic liquid crystals including nematic order and topological defects - singular points in the particle orientation field (see \cite{DIYS18} for a recent review).
Building on this established model, our active phase is controlled by a concentration parameter $\phi=1$ with an underlying orientation field captured in the continuum by a nematic tensor $\bmQ$, where the degree of nematic alignment is given by the largest eigenvalue of $\bmQ$ and the director is the corresponding eigenvector.
Both the concentration $\phi$ and $\bmQ$ vanish to zero outside the active phase.

The viscoelastic effects coming from polymer structures in the fluid are modelled in the continuum by using constitutive equations that describe the polymer stress contribution and its evolution (see, \textit{e.g.}, Refs.~\cite{ZLB12,SLP13,BU14,EL15,LA16,DD17,DNHE17,WWCLTMEO17}).
In our work, polymers are characterised by a conformation tensor $\bmC$ following the Oldroyd-B constitutive model \cite{BE94} and are suspended in the isotropic region ($\bmQ=\bm 0,\phi=0$). 
The Oldroyd-B model is one of the simplest and most widely-used polymer models since it features a stress contribution and relaxation rate linear with respect to $\bmC$, but nevertheless reproduces real viscoelastic phenomena like drag reduction in high-Reynolds fluid flows or elastic turbulence in low-Reynolds regimes \cite{BC18}. 
The Oldroyd-B model also corresponds to a kinetic model that describes the polymers as linearly-extensible dumbbells, but it does not prohibit infinite length. 
The dimensionless conformation tensor is in equilibrium if $\bmC=\bmI$, its trace $\trace{\bmC}$ characterizes the square of the polymer elongation, and the eigenvector corresponding to its largest eigenvalue gives the polymer orientation.

\subsection{Governing equations} 
The dynamical equations we solve are the incompressible Navier-Stokes equations for the fluid velocity $\bm u$, the Cahn-Hilliard equation for the concentration $\phi$, the Beris-Edwards model for the liquid crystal order parameter $\bmQ$, and the Oldroyd-B model for the polymer conformation tensor $\bmC$ \cite{DE86,BE94}:
\begin{eqnarray}
\rho (\partial_t \bmu + \bmu \bcdot \bnabla \bmu)
&=&\bnabla\bcdot\bPi, \quad (\bnabla \bcdot\bmu=\bm 0),
\label{eq:NSE}\\
\partial_t\phi+\bnabla\bcdot(\phi\bmu)
&=&\Gamma_\phi\bnabla^2\mu, 
\label{eq:CH}\\
\partial_t\bmQ+\bmu\bcdot\bnabla\bmQ 
&=&\bmS_Q+\Gamma_Q\bmH_Q
\label{eq:Q_eq},\\
\partial_t\bmC+\bmu\bcdot\bnabla\bmC 
&=&\bmS_C+\Gamma_C[\bmH_C\bmC+\bmC^\top\bmH_C^\top]+K_C\bnabla^2\bmC
\label{eq:C_eq},
\end{eqnarray}
where $\rho$ is the fluid density and $\bPi=-p\bmI+\bm\sigma$ denotes the total stress with $p$ the pressure, $\bmI$ the identity tensor and $\bm\sigma$ denoting deviatoric stresses which are listed below.
$\Gamma_\phi, \Gamma_Q$ and $\Gamma_C$ control the speed of relaxation of the order parameters to a free energy \Em{minimum} in their respective fields $\mu,\bmH_Q$ and $\bmH_C$. The corotational tensors
$\bmS_Q=(\xi\bmD+\bm\Omega)(\bmQ+\bmI/3)+(\bmQ+\bmI/3)(\xi\bmD-\bm\Omega)-2\xi(\bmQ+\bmI/3)\trace{\bmQ\bnabla\bmu}$ and $\bmS_C=\bmC\bm\Omega-\bm\Omega\bmC+\bmC\bmD+\bmD^\top\bmC^\top$ describe the rotational and elongational dynamics of $\bmQ$ and $\bmC$ in response to velocity gradients appearing via the vorticity $\bm\Omega$ and strain $\bmD$ tensors; $\xi$ is a tumbling parameter that determines the extent to which the nematic directors tumble in or align with the flow.
A diffusion term in \eqref{eq:C_eq} with coefficient $K_C$ is necessary to maintain numerical stability \cite{SB95}.

The explicit molecular fields and stresses can be derived from the local free energy $f=f_{Q}+f_{C}+f_{\phi}$, where the individual components are
\begin{eqnarray}
f_{Q}&=&A_Q\left[\dfrac{1}{2}\left(1-\dfrac{\eta(\phi)}{3}\right)\trace{\bmQ^2}-\dfrac{\eta(\phi)}{3}\trace{\bmQ^3}+\dfrac{\eta(\phi)}{4}\trace{\bmQ^2}^2\right]+\dfrac{K_Q}{2}(\bnabla\bmQ)^2+L_0(\bnabla\phi\bcdot\bmQ\bcdot\bnabla\phi), \label{eq:fe_Q}\\
f_{C}&=&\dfrac{A_C}{2}\Em{(1-\phi)}(\trace{\bmC-\bmI}-\ln\operatorname{det}\bmC), \label{eq:fe_C}\\
f_\phi&=&\dfrac{A_\phi}{2}\phi^2(1-\phi)^2 +\dfrac{1}{2}K_\phi(\bnabla\phi)^2. \label{eq:fe_phi}
\end{eqnarray}
$A_Q$, $A_C$, and $A_\phi$ control the attractivity of the equilibrium configurations, $K_Q$ and $K_\phi$ impose the strength of the energy penalties associated with gradients, and $L_0$ enforces parallel or perpendicular nematic interface anchoring. 
The first term in $f_{Q}$ is the Landau--de Gennes expression for the free energy of lyotropic nematic liquid crystals, controlled by a phase-dependent temperature $\eta(\phi)$ with a critical value $\eta_{\rm crit}=2.7$ \Em{below} which the nematic ordering dissolves; here, we adopt $\eta(\phi)=\eta_{\rm crit}+(\phi-\phi_{\rm boun})/2$ with a boundary defined by $\phi_{\rm boun}=0.5$~\cite{SMY06}. 
\Em{The contribution of the viscoelastic phase appears in \eqref{eq:fe_C}, as given by the Oldroyd-B model.}
The first expression in \eqref{eq:fe_phi} favors either  of the values $\phi=0$ or $\phi=1$ in the concentration.
The molecular fields are then written as functional derivatives of $f$ \cite{HMBMRFC15}:
\begin{eqnarray}
\bmH_Q
&=&-A_Q\left[\left(1-\dfrac{\eta}{3}\right)\bmQ-\eta\bmQ^2+\eta\bmQ^3\right]-\dfrac{\bmI}{3}A_Q\phi^2\eta\trace{\bmQ^2} +K_Q\bnabla^2\bmQ-L_0\bnabla\phi\bnabla\phi^\top, \label{eq:field_Q}\\
\bmH_C&=&
-\dfrac{A_C}{2}\Em{(1-\phi)}(\bmI-\bmC^{-1}), \label{eq:field_C}\\
\mu
&=&A_\phi\phi(1-\phi)(1-2\phi)-K_\phi\bnabla^2\phi+\dfrac{A_Q}{2}\left(-\dfrac{1}{6}\trace{\bmQ^2}-\dfrac{1}{3}\trace{\bmQ^3}+\dfrac{1}{4}\trace{\bmQ^2}^2\right)\nn\\ &&\Em{-\dfrac{A_C}{2}(\trace{\bmC-\bmI}-\ln\operatorname{det}\bmC)}-2L_0\left[(\bnabla\bcdot\bmQ)\bcdot\bnabla\phi+\bmQ\bm:(\bnabla\otimes\bnabla\phi)\right], \label{eq:field_phi}
\end{eqnarray}
where $\bm{\mathrm{A:B}}=\sum_{i,j}A_{ij}B_{ji}$ and $\bm a\otimes\bm b=a_ib_j$.

Adopting the usual approach to incorporate activity in the total stress tensor \cite{DIYS18}, we write the total deviatoric stress $\bm\sigma$ in \eqref{eq:NSE} as the sum of the following components:
\begin{eqnarray}
\bm\sigma_{\rm active}&=&-\zeta\phi\bmQ, \label{eq:astress}\\
\bm\sigma_{\rm capillary}
&=&(f-\phi\mu)\bmI-K_\phi(\bnabla\phi)\otimes(\bnabla\phi)-2L_0(\bnabla\phi\bcdot\bmQ\bcdot\bnabla\phi), \label{eq:cstress}\\
\bm\sigma_{\rm elastic (nematic)}
&=&-\xi\bmH_Q(\bmQ+\dfrac{1}{3}\bmI)-\xi(\bmQ+\dfrac{1}{3}\bmI)\bmH_Q+2\xi\trace{ \bmQ\bmH_Q}(\bmQ+\dfrac{1}{3}\bmI)\nn\\
&&+\bmQ\bmH_Q-\bmH_Q\bmQ-K_Q(\bnabla\bmQ)\bm:(\bnabla\bmQ), \label{eq:estress}\\
\bm\sigma_{\rm polymer}&=&-2\bmH_C^\top\bmC^\top, \label{eq:pstress}\\
\bm\sigma_{\rm viscous}&=&2\nu\bmD. \label{eq:vstress}
\end{eqnarray}
$\nu$ is the dynamic solvent viscosity and $\zeta$ measures the level of activity with positive (negative) values corresponding to extensile (contractile) stresses. \Em{We note that while most studies of active matter keep the activity fixed, we will exploit the freedom to allow $\zeta$ to depend on its location in the active phase.  }\\

Note that whereas both $\bmQ$ and $\bmC$ are defined throughout the domain, $\bmQ$ vanishes to $\bm 0$ in the viscoelastic region via \eqref{eq:fe_Q} while $\bmC$ has to be numerically forced to remain at its equilibrium value $\bmI$ in the active region.
Lastly, we note that this formulation assumes that the active matter and the polymers are thermodynamically independent (no coupling between $\bmQ$ and $\bmC$ in the free energies) and only interact via the velocity in \eqref{eq:NSE}.

\subsection{Simulation details}

Equations \eqref{eq:NSE}--\eqref{eq:C_eq} are solved by using a hybrid lattice-Boltzmann (LB) method \cite{MOCY07,BTY14}. 
The conformation tensor $\bmC$ was evolved by a finite-difference scheme and remained positive-definite in the simulations.
All simulations are performed in a two-dimensional periodic box with length $L=100$. 
Throughout the study, we assume $\rho=1,~p=0.25$, and $\nu=2/3$. 
The liquid crystal parameters remain fixed:
$A_Q=0.5, \xi=0.3, K_Q=0.02, \Gamma_Q=0.25, K_C=0.01$.
The parameters controlling the interface vary between simulations.
For cell division, the parameters are $A_\phi=0.25, K_\phi=0.04, \Gamma_\phi=0.08, L_0=0$.
For the motile cell, the parameters are $A_\phi=0.07, K_\phi=0.04, \Gamma_\phi=0.08, L_0=-0.05$.
For interfacial instabilities, the parameters are as  $A_\phi=0.2, K_\phi=0.03, \Gamma_\phi=0.01, L_0=0$. 
\Em{These parameters were chosen to lie within the range used in previous studies using active nematics to model biological systems~\cite{DTSLLY15,DTY16}; the relation to physical units will be discussed for each simulation in the next section. }\\

The fluid is initially at rest, $\bmu=\bm 0$, unless otherwise stated and the polymers, if present, are all initialised in equilibrium ($\bmC=\bmI$). 
The initial conditions of $\bmQ$ when $\phi=1$ differ for each simulation and will be described separately.

It is more intuitive to write the dynamical equation for $\bmC$ and the polymer stress $\bm\sigma_{\rm polymer}$ in terms of physical parameters that describe the polymeric fluid. Here we denote the polymer relaxation time as $\tau_C$ and the polymer viscosity as $\nu_C$.
Polymers with larger $\tau_C$ are more elastic and require a longer time to relax back to their equilibrium length $\bmI$.
Meanwhile, the polymer viscosity $\nu_C$ describes how much the polymers contribute to the total viscosity of the system. Since the increase in viscosity is proportional to the number of polymers that are introduced into the fluid, the value of $\nu_C$ may as well be interpreted as the polymer concentration of the fluid.
Thanks to the established relations $\Gamma_C=\nu_C^{-1}$ and $A_C=\nu_C/\tau_C$ \cite{BE94}, the right-hand side in \eqref{eq:C_eq} simplifies to $\bmS_C-(\bmC-\bmI)/\tau_C+K_C\bnabla^2\bmC$, and the polymer stress \eqref{eq:pstress} reads:
\begin{equation}
\bm\sigma_{\rm polymer}=\frac{\nu_C}{\tau_C}(1-\phi)(\bmC-\bmI).
\label{eq:polystress_simp}
\end{equation}
Investigating a surrounding viscoelastic environment using this polymer model is then reduced to understanding the effect of varying the polymer relaxation time $\tau_C$ and the polymer viscosity $\nu_C$ independently.
In each of the simulations that follow, we then either vary $\tau_C$ and keep $\nu_C=1$ fixed ($\tau_C$-case), or vary $\nu_C$ and keep $\tau_C=1000$ fixed ($\nu_C$-case).

\section{Results and discussion}
To illustrate the effectiveness of the model in capturing the interaction between active matter and a viscoelastic environment we first study two important examples of cell dynamics in a viscoelastic medium: 
A. cell division, and 
B. cell motility.
The effects of both polymer relaxation time $\tau_C$ and polymer viscosity $\nu_C$ are then discussed.
Lastly, we generalise the framework to show how the generic formation of instabilities at an active matter/fluid interface are mitigated by viscoelasticity.

\subsection{Cell division in a viscoelastic medium}
From a mechanical perspective, cell division begins with a phase of swelling characterised by an increase in volume (interphase).  
Cells then exert protrusive forces on their environment (anaphase) to open up space for elongation along their division direction, the mitotic axis. 
Only if the cell manages to elongate sufficiently can cell division occur.
Two main physical mechanisms contribute to the protrusive forces \cite{NC18}: 
the elongation of the microtubule assembly along the mitotic axis, and 
the contraction of the cytokinetic actymyosin ring at the equator, which occurs perpendicular to the elongation axis and cleaves the cell.
While both sources of stress are indispensable, elongation accounts for $\sim 80\%$ of the initial protrusive stress (pre-anaphase) while ring contraction accounts for $\sim 88\%$ of the stress in the latter stages (post-anaphase). 

\begin{figure}[t]
\centering
\includegraphics[width=0.75\textwidth]{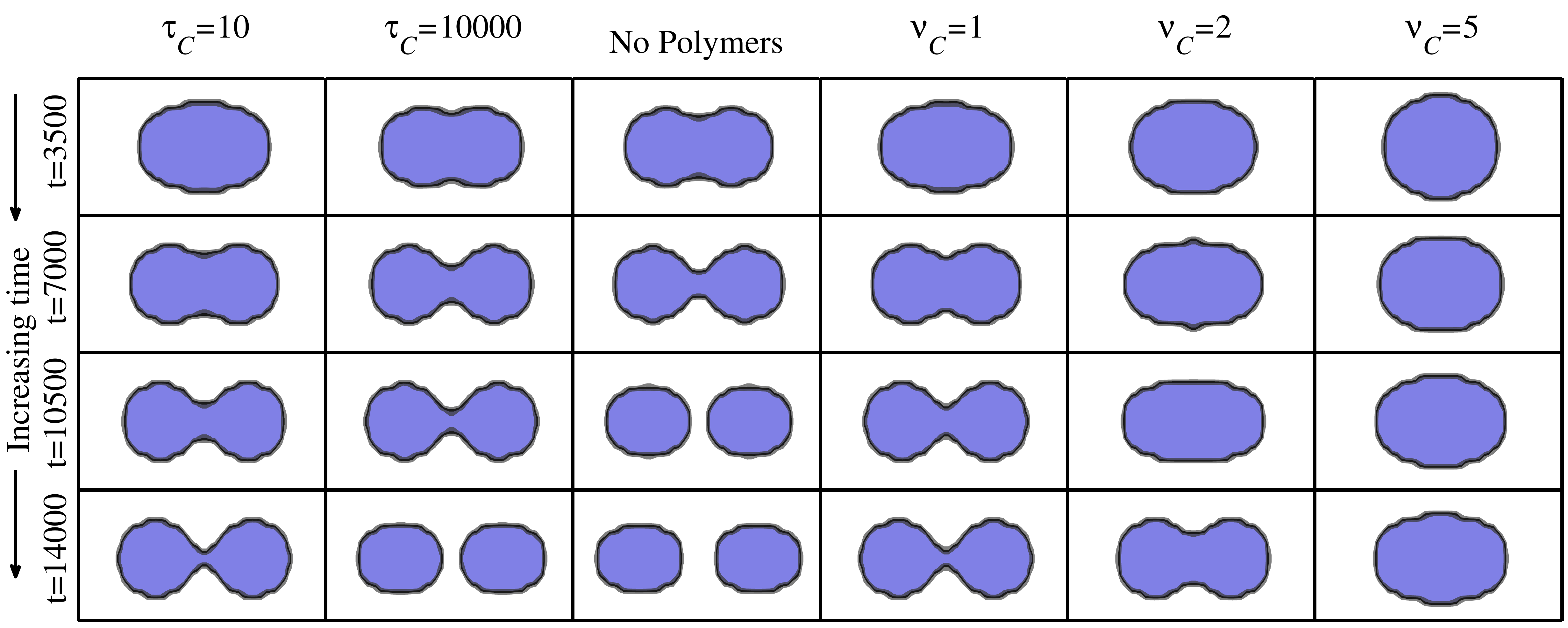}
\caption{Simulation snapshots of a cell undergoing division for the $\tau_C$-case (first two columns), the no-polymer case (third column), and the $\nu_C$-case (last three columns)}
\label{fig:mitosisevol}
\end{figure}

A recent experimental study reported that mammalian cells confined in stiff hydrogels are hard-pressed to divide and that the probability of cell division decreases inversely with hydrogel stiffness \cite{NC18}. 
Moreover, excessive delays in cell division due to the presence of stiff hydrogels may result in programmed cell death \cite{CHAC15, ASRZC18,NC18}.
Motivated by this experiment, we first investigate the effect of a viscoelastic medium on the cell-division process.
In our simulations we assume that the cell has finished the mitotic swelling and is about to elongate (pre-anaphase). 
The cell is modelled as a circular region of active matter, in which the nematic director field represents the microtubules within the cell that actively generate mechanical stresses to elongate the cell.
Such an approach has been previously applied to model a dividing cell as an active nematic region in an isotropic fluid medium~\cite{GD14,Marchetti17}.
The initial cell is a circle with radius 10 LB units, with all the directors aligned to the horizontal axis (see Fig.~\ref{fig:mitosisevol}). 
The elongational effect of the microtubules manifests through an extensile nematic activity $\zeta_{\rm spindle}=\zeta=0.006$ throughout the cell, while the cleaving effect of the cytokinetic ring is captured by an intensified local activity $\zeta_{\rm ring}=\zeta_{\rm spindle}+\zeta/4$ in a circular region of radius $r_{\rm ring}=4$ LB units at the centre of the cell.
These sources of stress are modulated such that contraction becomes more significant when the perimeter of cell has increased by at least 30\%: 
$\zeta_{\rm spindle}=0.7\zeta$ and $\zeta_{\rm ring}=\zeta_{\rm spindle}+3\zeta$.
In accordance with the experiments~\cite{NC18}, we verified that cell division will fail in the absence of either stress.

\begin{figure}[t]
\centering
\includegraphics[width=0.45\textwidth]{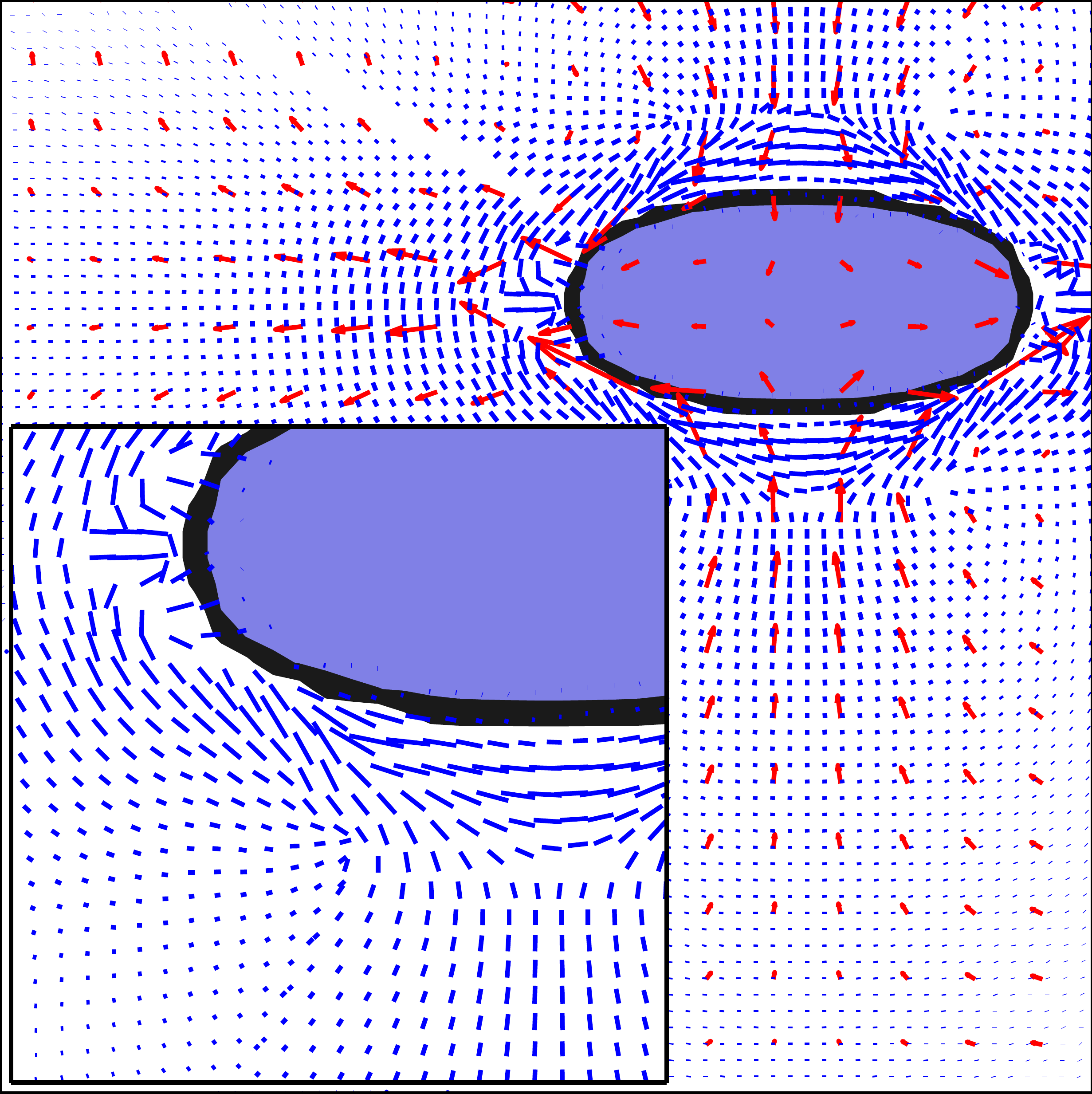}
\caption{Typical flow field of a cell elongating during the initial steps of cell division. Velocity vectors are in red, and polymer deformations \Em{$\bmC-\bmI$} are in blue. Inset: Zoom of the lower left part of the cell, showing polymers aligning with the stretching direction. Polymer deformations and velocities are rescaled by 150\% for visibility.}
\label{fig:mitosis_flow}
\end{figure}

\begin{figure}[t]
\centering
(a)\includegraphics[width=0.46\textwidth]{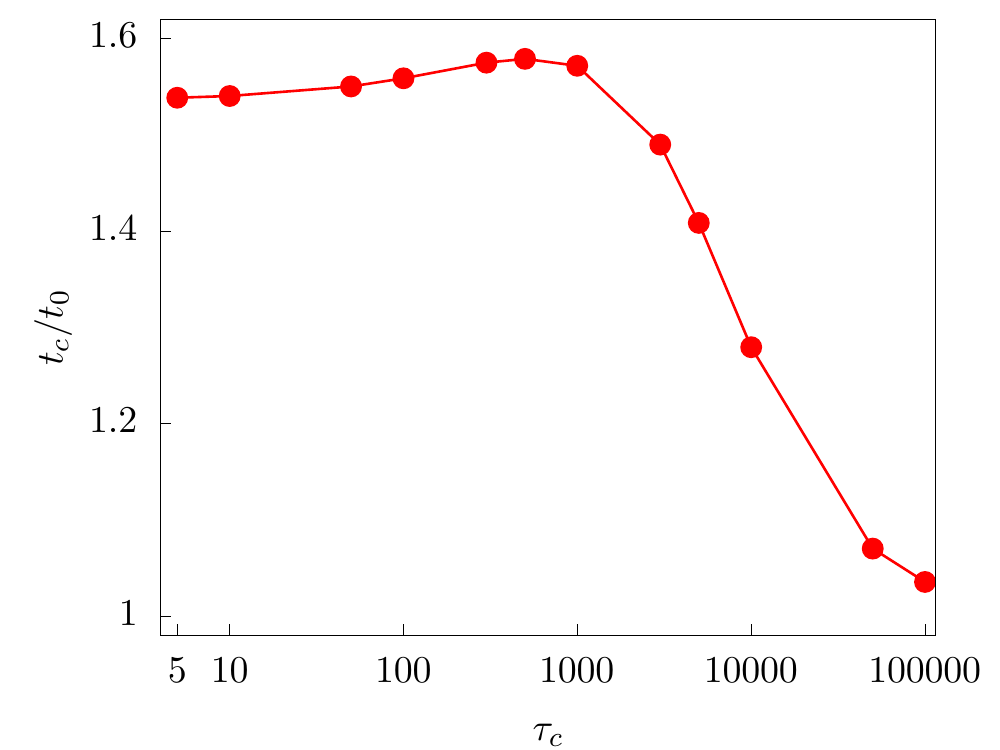}
(b)\includegraphics[width=0.46\textwidth]{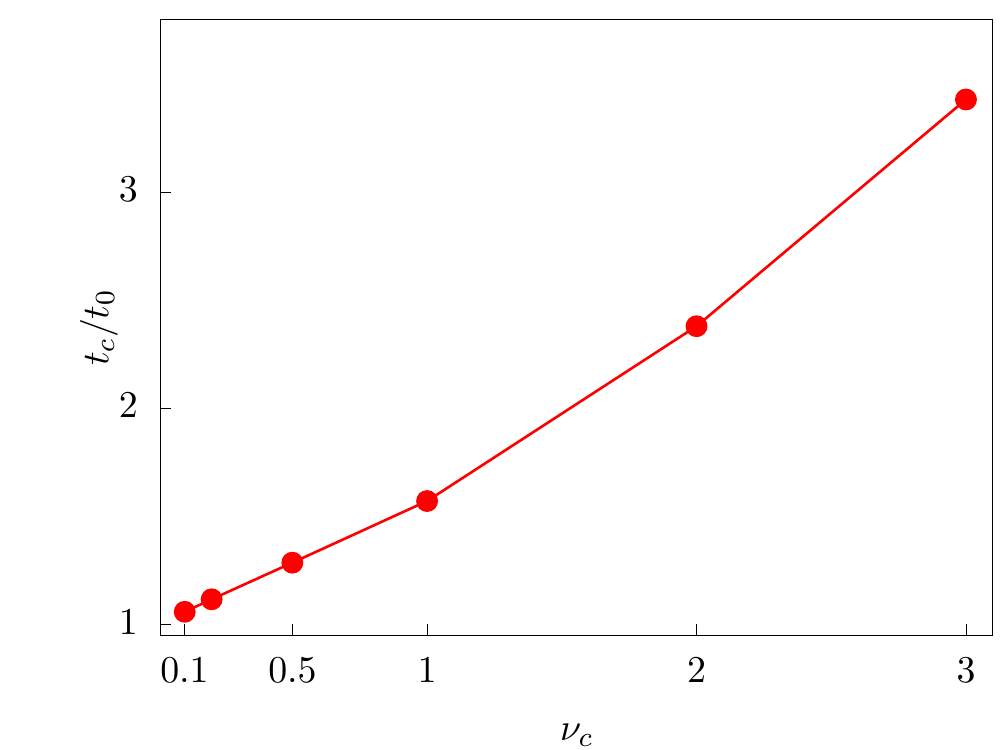}
\caption{Total time for cell division in a viscoelastic fluid $t_c$, normalized by the total time to divide in the no-polymer case $t_0$, for (a) the $\tau_C$-case and (b)  the $\nu_C$-case.}
\label{fig:mitosistime}
\end{figure}

In the absence of polymers, the circular cell completes division after around $10^4$ LB time steps (Fig.~\ref{fig:mitosisevol}, 3rd column).
When the cell is placed in a polymeric fluid, division is generally delayed and can even be almost completely suppressed if the polymer viscosity $\nu_C$ is large, as seen in Fig.~\ref{fig:mitosisevol}. 
At early times the velocity field exhibits a straining flow (Fig.~\ref{fig:mitosis_flow}), with flow fields qualitatively comparable to those in the experiments.
The polymers near the cell interface stretch and align in the stretching direction of the flow (see inset of Fig.~\ref{fig:mitosis_flow}).
The contractile nature of the polymer stress ($\bm\sigma_{\rm poly}\propto \bmC$) hence reduces the magnitude of the velocity fields, effectively delaying cell division.
This delay is quantified by comparing the amount of time required to perform cell division in a polymeric fluid $t_c$ to that in a Newtonian fluid $t_0$ (Fig.~\ref{fig:mitosistime}).
Note that $t_c$ is considerably larger than $t_0$ for $\tau_C<10^4$ with little distinction between low ($\tau_C\sim10$) and intermediate ($\tau_C\sim10^3$) values. 
Meanwhile, extremely elastic polymers ($\tau_C>10^4$) allow the cell to perform division at a speed not far from the no-polymer case.
In the $\nu_C$-case, however, increasing polymer viscosity $\nu_C$ results in a nearly-linear increase in $t_C$, with cell division failing to occur for $\nu_C\geq4$. 

The retardation effect of the polymers on cell division provides validation that the model has potential to capture cellular dynamics in viscoelastic environments.
The mechanism for the generic impacts of relaxation time $\tau_C$ or polymer viscosity $\nu_C$ on active matter dynamics will be explained in the next section, where we consider the self-propulsion of a model cell within a viscoelastic surrounding.
We will also see that the non-monotonic dependence on the polymer relaxation time and nearly-linear response to changing polymer viscosity are overarching themes when active matter interacts with a viscoelastic medium. 

It is instructive to note that for the cells undergoing division, we can map our cell diameter of 20 LB grid lengths to experimental cells with diameter of $\approx 15 \mu$m. 
The elastic gels used in the experiment~\cite{NC18} have a  relaxation time of $10^4$ seconds, and if mapped to $\tau_C=10^3$ LB time steps, an approximate splitting time of $10^3$ LB time steps translates to 26 hours in real time (not including initial swelling), which is within estimates of the 24- to 38-hour doubling time of the MDA-MB-231 cells used in the experiment.
The maximum (principal) active stress during elongation in simulations is 0.0025 LB force units corresponds to 2.0 kPa protrusive stress in experiments.

\subsection{Cell motility in a viscoelastic medium}
\begin{figure}[t]
\centering
\includegraphics[width=0.75\textwidth]{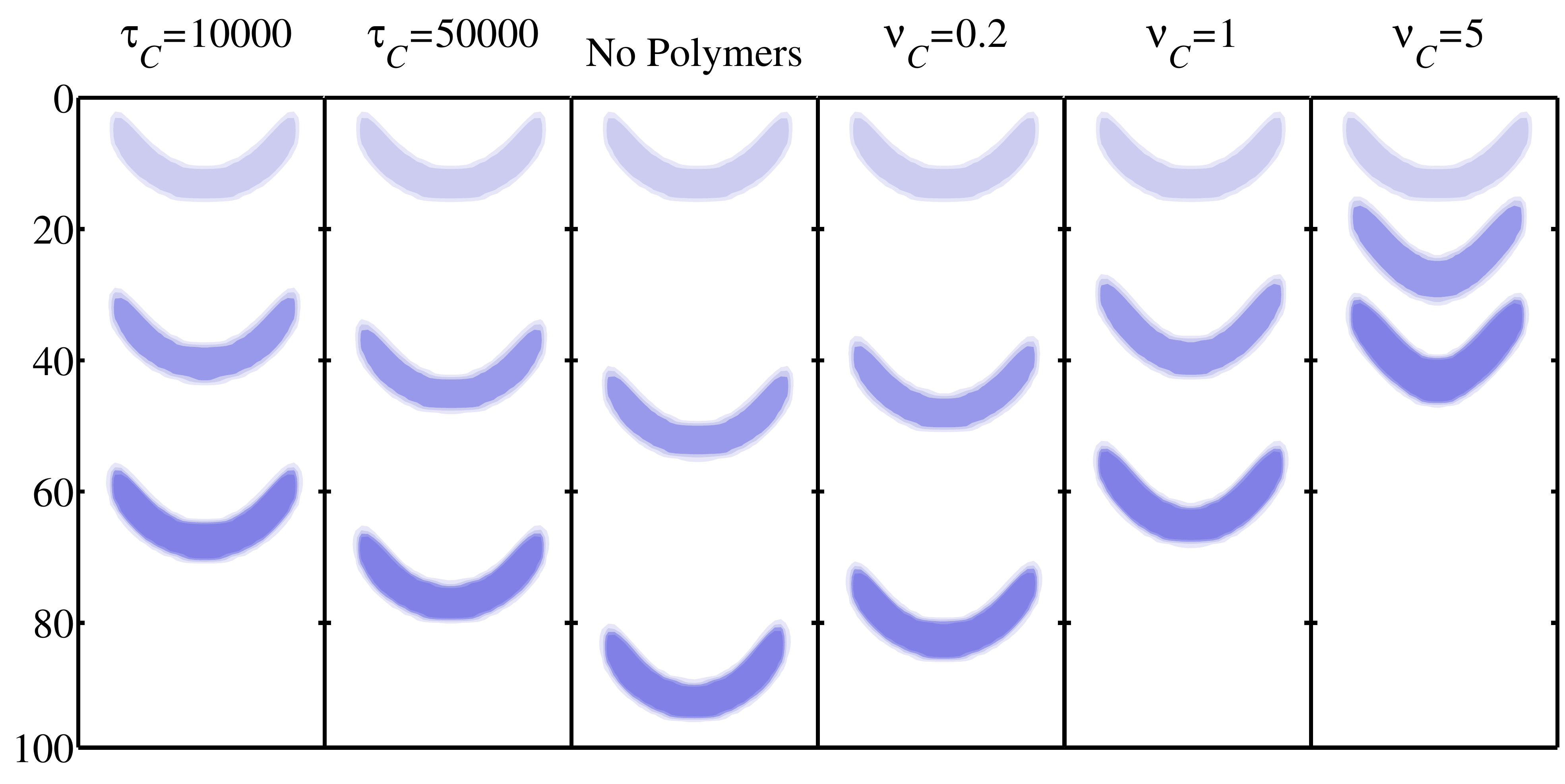}
\caption{Simulation snapshots of a moving model keratocyte cell for the $\tau_C$-case (first two columns), the no-polymer case (third column), and the $\nu_C$-case (last three columns).  Snapshots were taken at times $t=0$ (top, very light blue), $t=3.5\times10^5$ (middle, light blue) and $t=7\times10^5$ (bottom, blue) after initialisation.}
\label{fig:crescentmove}
\end{figure}

To understand the physical mechanisms at work in active-viscoelastic systems, next we examine a cell that is translating in mucous fluid, and with little or no observable changes in its morphology.
One well-established choice to study motile cells is crawling fan-shaped keratocytes with an actin-induced motion and substrate adhesion \cite{KT08,A16}.
The model proposed here, Eqs.~\eqref{eq:NSE}--\eqref{eq:C_eq}, can reproduce moving cell dynamics morphologically reminiscent of keratocyte cells by coupling perpendicular nematic anchoring at the boundary with contractile active stresses ($\zeta=-0.005<0$). 
Given suitable parameters, a circular cell with a radius of 8 LB units elongates, bends, and equilibrates into a crescent-shaped cell that translates in the direction perpendicular to its long axis (see Fig.~\ref{fig:crescentmove})~\cite{TMC12,GD14}\Em{; the  symmetry-breaking is a consequence of the well-known bend instability in active nematics~\cite{MJRLPRA13,R10}.}
To ensure that the differences between the cases do not arise due to the initial elongation of the cell, we consider a cell that has equilibrated into a crescent and is already translating in an isotropic fluid (after $3\times 10^5$ LB time steps), and simply incorporate the polymeric component if needed.
In this case mapping 1 LB grid length to 1.5$\mu$m and 1 LB time step to $6.4\times 10^{-4}$s gives a simulated cell velocity ($\approx 85.5\times10^{-6}$ LB velocity units) that corresponds to the $\approx 0.2 \mu$m/s mean velocity of fish keratocyte cells with medium adhesive strength \cite{BLKMT11}. \Em{Reynolds numbers are small, typically $O(10^{-2}$)-$O(10^{-3})$.}

\begin{figure}[t]
\centering
(a)\includegraphics[width=0.46\textwidth]{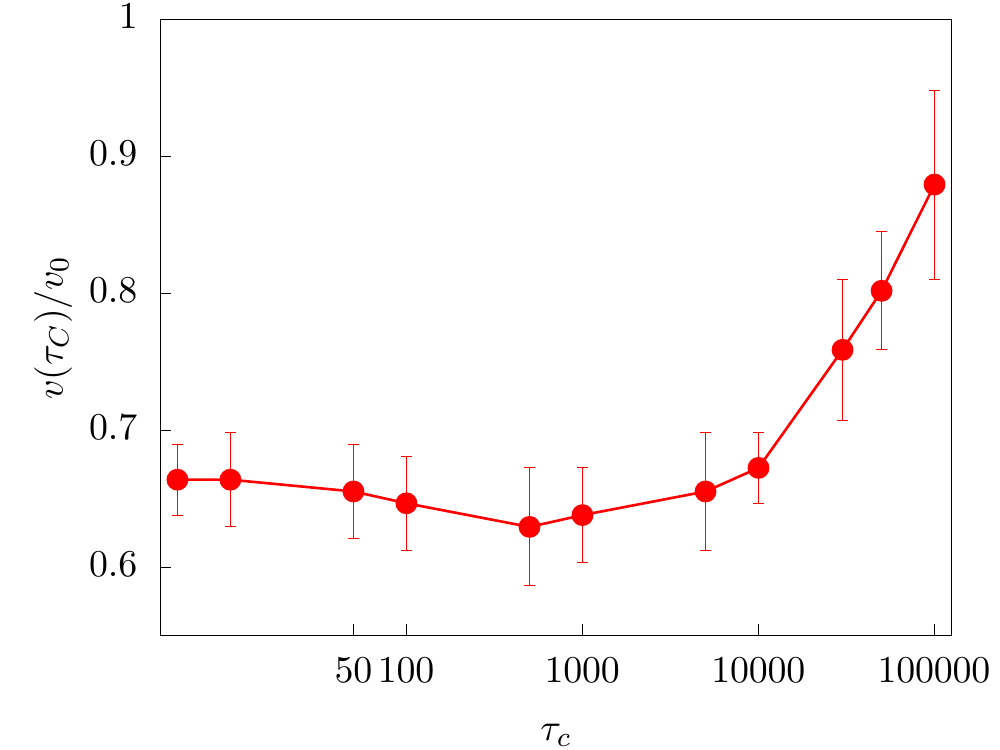}
(b)\includegraphics[width=0.46\textwidth]{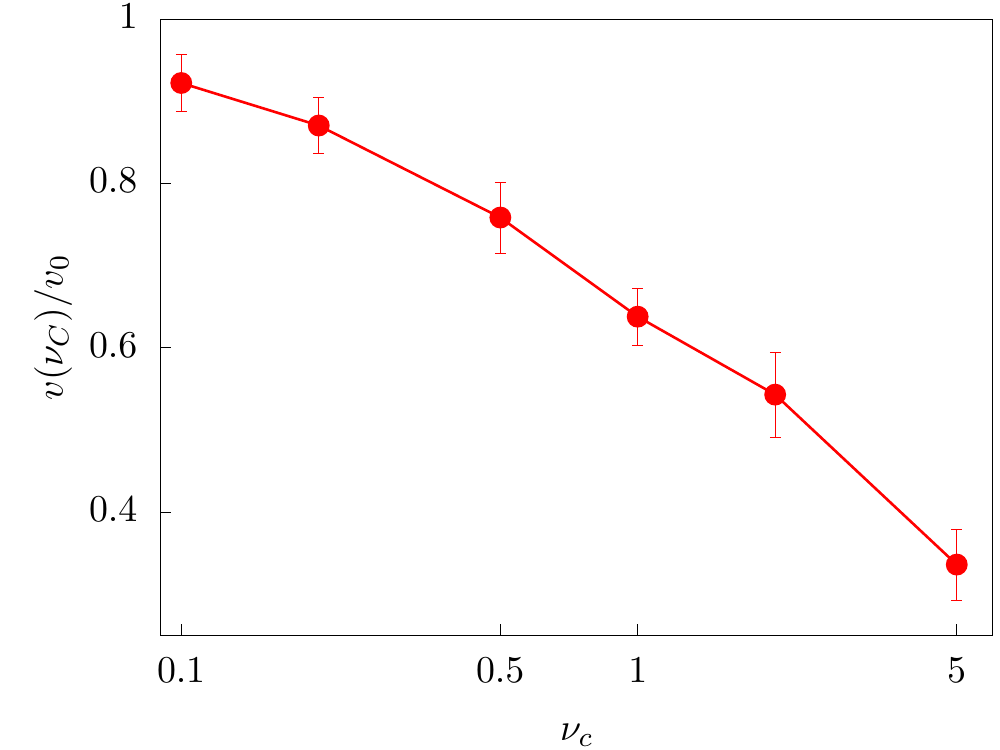}
\caption{Mean vertical velocity of the centre of mass of the model keratocyte cell in a viscoelastic fluid $v$, normalized by the mean velocity in a Newtonian fluid $v_0$ for (a) the $\tau_C$-case and (b)  the $\nu_C$-case. Error bars represent 1 standard deviation of averaged velocities sampled 60 times over $1.2\times10^6$ LB time steps.}
\label{fig:crescentspeed}
\end{figure}

Figure~\ref{fig:crescentmove} compares snapshots of the motile cell moving through an isotropic or polymeric fluid, showing that the polymeric fluid clearly slows down cell movement. 
With or without polymers in the surrounding fluid the crescent-shaped cell moves at almost constant velocity.
The velocity of the centre of mass of the cell is reported in Fig.~\ref{fig:crescentspeed} as a function of the viscoelastic parameters.
In the $\tau_C$-case (fixed polymer viscosity and varying polymer relaxation time), the velocity displays values nearly independent of the relaxation time $\tau_C$ for low to moderate values of $\tau_C$, but significantly increases for $\tau_C$ above $10^4$.
\Em{
In the $\nu_C$-case (fixed polymer relaxation time and varying polymer viscosity), the velocity of the cell
} monotonically decreases as the polymer viscosity $\nu_C$ increases, similar to the damping effect seen in the dividing cell.

\begin{figure}[t]
\centering
(a)\includegraphics[width=0.46\textwidth]{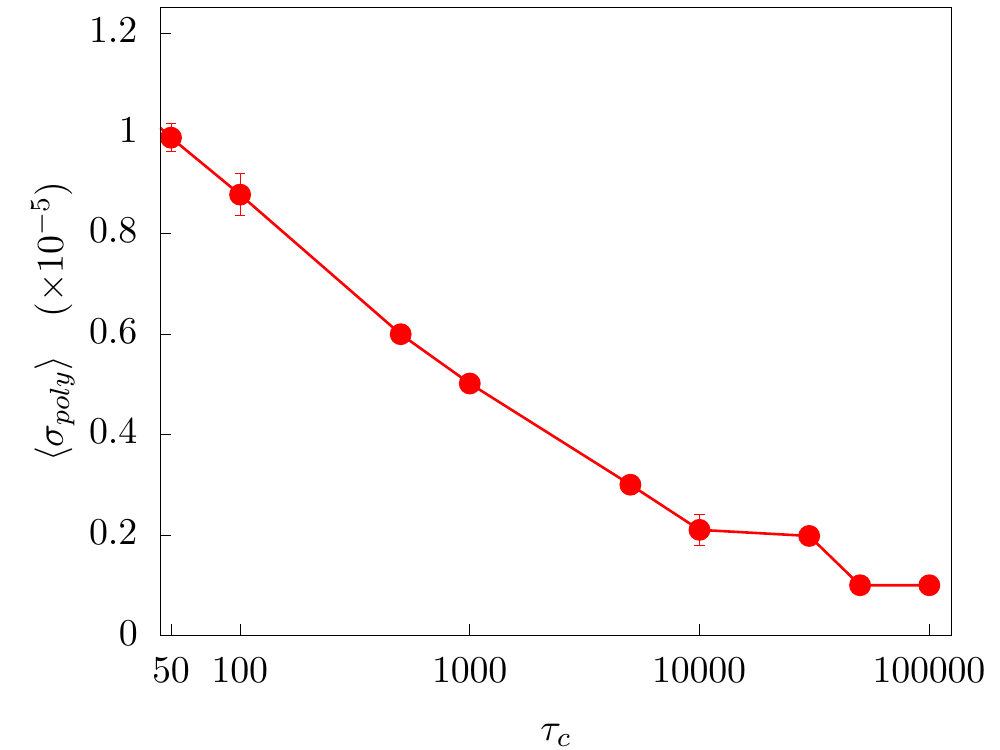}
(b)\includegraphics[width=0.46\textwidth]{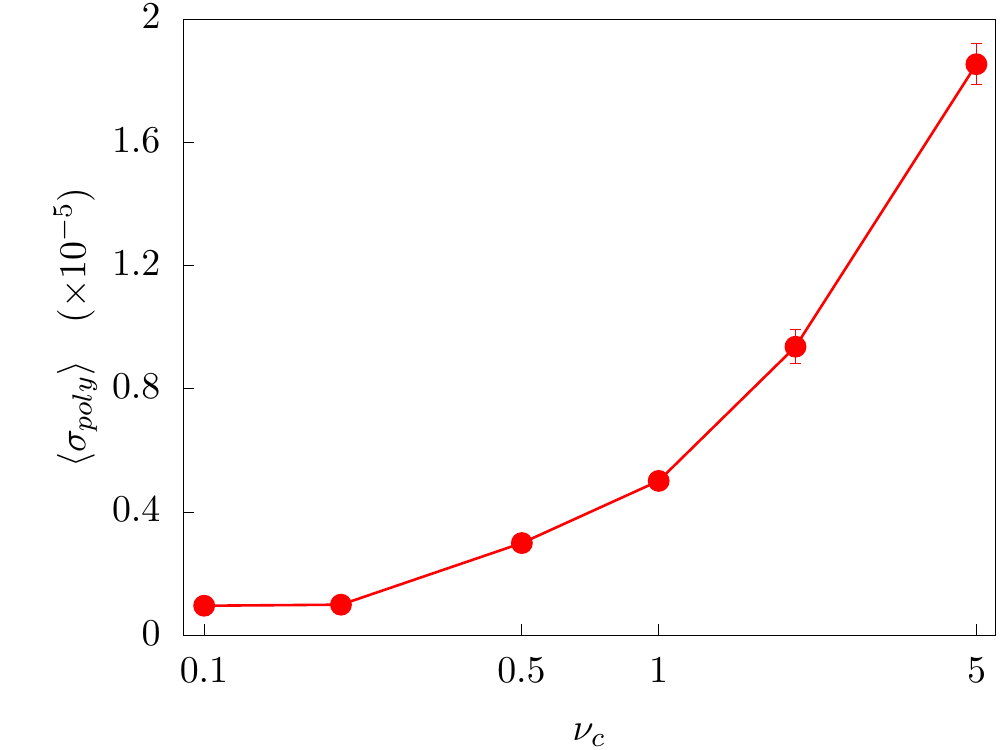}
\caption{Mean magnitude of the principal polymer stress for a model keratocyte cell propelling through a viscoelastic fluid for (a) the $\tau_C$-case and (b)  the $\nu_C$-case.
Error bars represent 1 standard deviation calculated as in Fig.~\ref{fig:crescentspeed}.}
\label{fig:pstress_crescent}
\end{figure}

We have seen in both cell motility and cell division studies that polymers slow down active matter dynamics, 
\textit{i.e.} the polymeric stress generated counteracts the activity-induced fluid velocity.
The strength of the polymeric stress is controlled by the polymer viscosity and relaxation time: $\bm\sigma_{\rm polymer}\propto(\bmC-\bmI)\nu_{C}/\tau_C$.
The effect of $\nu_C$ and $\tau_C$ on the polymer stress is evident for the moving cell since the cell does not deform and $\bm\sigma_{\rm polymer}$ fluctuates very little (Fig.~\ref{fig:pstress_crescent}).
Polymer stress on one hand is diminished as polymer relaxation time increases, with very large $\tau_C$ approaching no-polymer cases.
On the other hand, polymer stress is magnified as we increase polymer viscosity, where motion and deformations can even be arrested for exceedingly large $\nu_C$.

We draw a contrast between the simulations described above and what is observed in other studies, \textit{e.g.} \cite{HMBMRFC15,HCF16}, wherein using a large value of $\tau_C$ resulted in significant polymer feedback into the flow. 
In these reports the local Deborah number $De=\tau_C\sqrt{(\bmD\bm:\bmD)/2}$ exceeds the critical value $De>1/2$. 
Systems with $De$ above this value are characterised by polymers that stretch significantly which results in significant polymer stress that could bring the system into an elastic-turbulent flow \cite{GS01}.\\

\begin{figure}[t]
\centering
\includegraphics[width=0.46\textwidth]{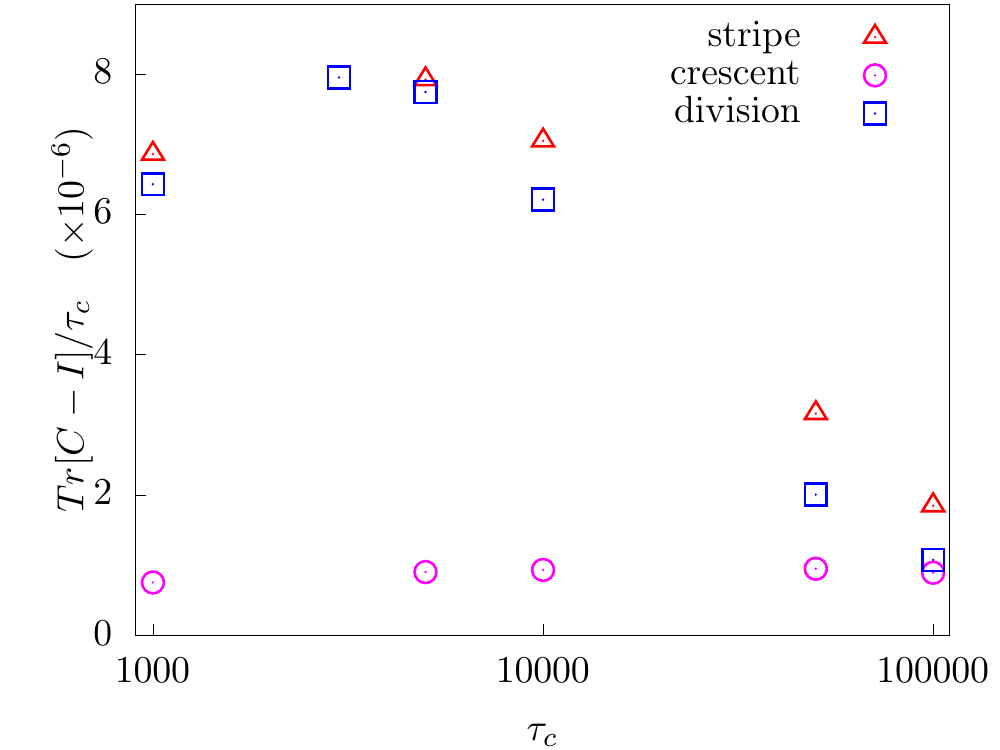}
\caption{Average polymer extension around the active phase, quantified via averaging $\trace{\bmC-\bmI}$ over time and space, divided by the polymer relaxation time $\tau_C$, as a function of $\tau_C$.}
\label{fig:lengths}
\end{figure}

\Em{In our simulations, however, since the nematic tensor $\bmQ$ is zero outside the cell, the driving active force ($-\zeta{\bf \nabla}\cdot\bmQ$) acts only within the cell or at its interface. This results in low strain rates within the polymer phase and, consequently, in small Deborah numbers.}
\Em{Indeed, figure~}\ref{fig:lengths} shows that $\trace{\bmC-\bmI}/\tau_C$ remains small even for more elastic polymers (larger values of $\tau_C$); polymer elongation in the simulations of the motile cell is smaller relative to the other simulations because the stationary shape of the active phase produces minimal velocity gradients and results in lower $De$ numbers.
While polymer elongation in all simulations still grows with $\tau_C$, it increases at a rate slower than $\tau_C$.
Polymer deviations $\bmC-\bmI$ therefore remain small compared to $\tau_C$ and $\bm\sigma_{\rm polymer}$ simply scales directly with $\nu_C$ and inversely to $\tau_C$.
Most importantly, we only observe damping effects and no chaotic flows arise within the polymeric phase.

\subsection{Interfacial instabilities}
\begin{figure}[t]
\centering
\includegraphics[width=0.75\textwidth]{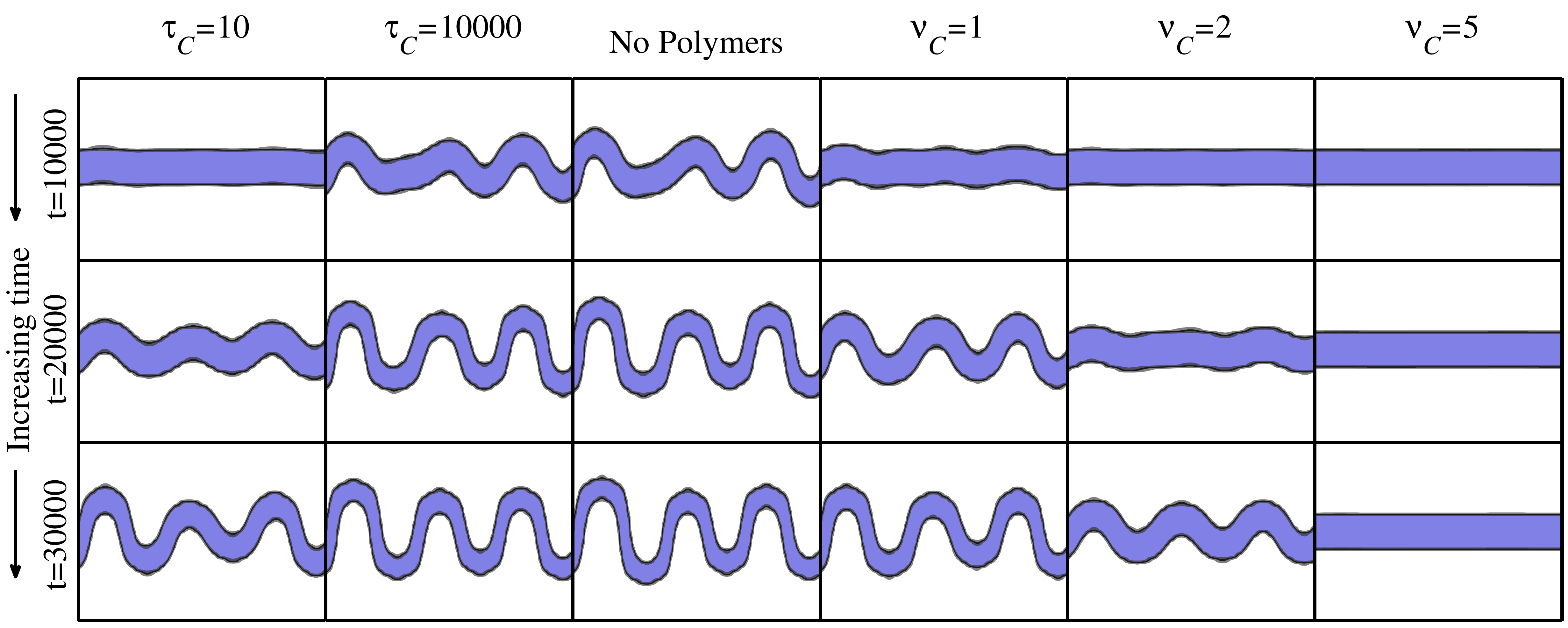}
\caption{Simulation snapshots of an active stripe developing deformations for the $\tau_C$-case (first two columns), the no-polymer case (third column), and the $\nu_C$-case (last three columns).}
\label{fig:evolution}
\end{figure}

Another way in which polymers can have an effect on the system is through the morphology of the active matter.
To highlight the polymer effect on the deformation of active interfaces, we next simulate a stripe of active nematics (width = 8 LB units) within a viscoelastic medium (see Fig.~\ref{fig:evolution}). 
The advantage of this setup is that it allows for directly assessing the impact of viscoelastic surrounding on the activity-induced instabilities.

It is well-established that bulk extensile (contractile) active nematics are unstable to bend (splay) deformations~\cite{SR02}. 
Therefore, in the absence of viscoelasticity, the activity-induced hydrodynamic instabilities in an extensile system lead to a bend deformation in an initially uniform director field. 
The bending within the stripe is accompanied by a bending of the interface, resulting in a wave-like deformation of the stripe with a well-defined wave length that depends on the activity~\cite{BTY14}.
In our simulation deformations are conspicuous at around 5000 LB time steps, after which they are amplified by induced fluid velocities (Fig.~\ref{fig:evolution}, 3rd column). 
The wavy stripe achieves and maintains a wavy configuration as shown in the bottom rows for long periods of time. 
Longer simulations indicate that this is not the final steady state, but we are only interested in the formation of initial instabilities.

\begin{figure}[t]
\centering
(a)\includegraphics[width=0.46\textwidth]{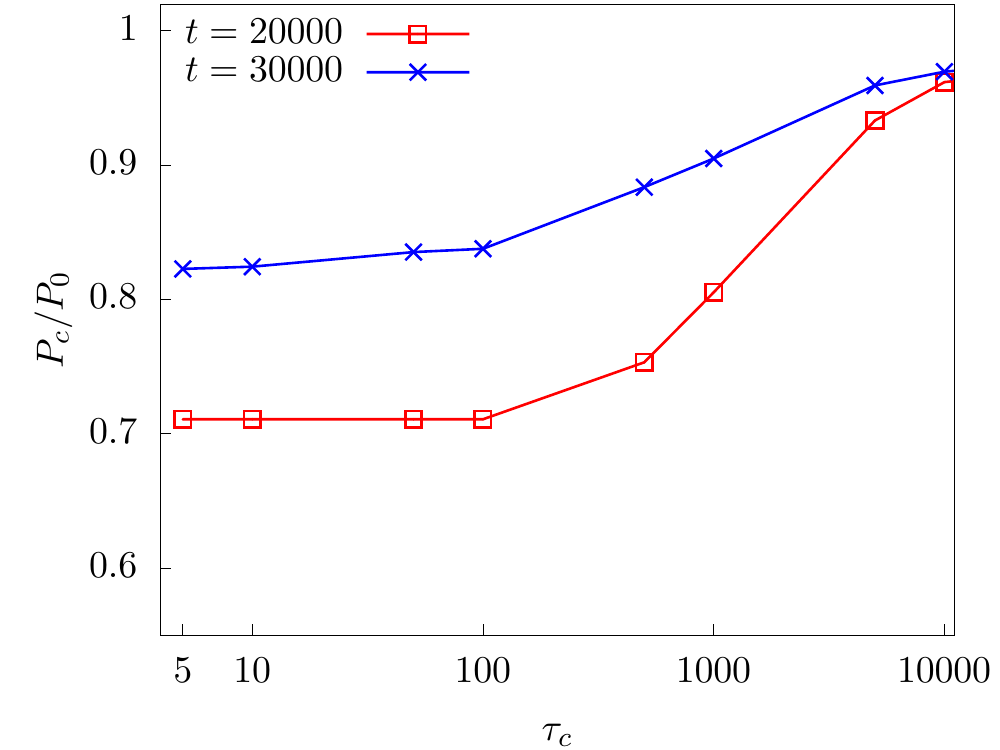}
(b)\includegraphics[width=0.46\textwidth]{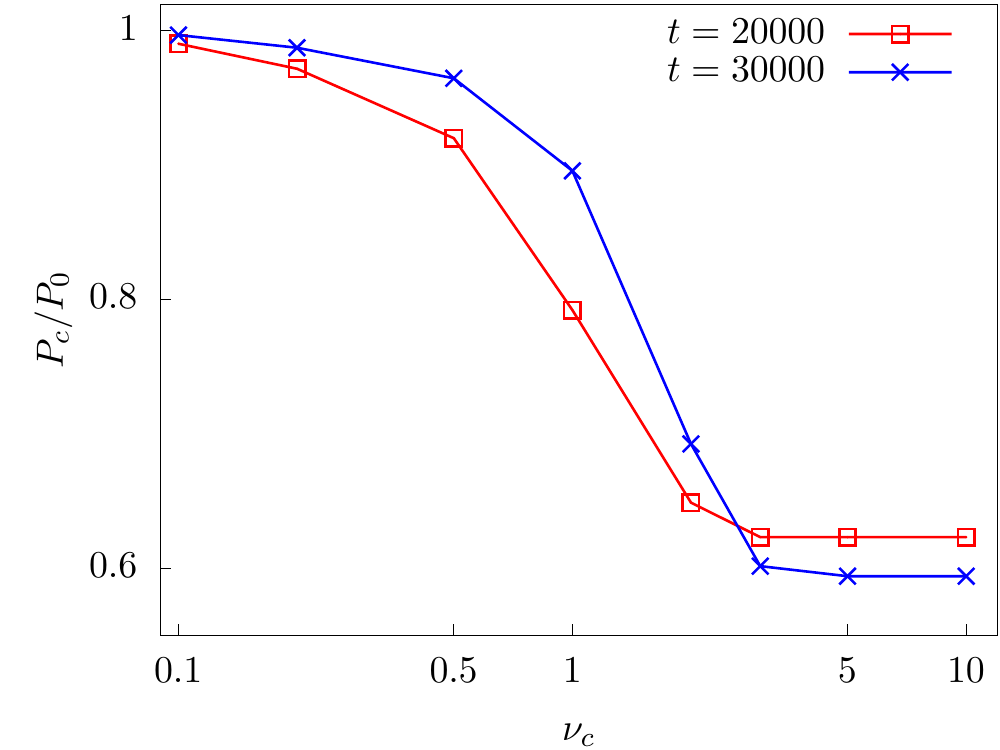}
\caption{Perimeter of the active stripe in a viscoelastic fluid $P_C$, normalized by the perimeter in the no-polymer case  $P_0$ at the same time for (a) the $\tau_C$-case and (b)  the $\nu_C$-case. Different curves refer to different points in time $t$.}
\label{fig:peri}
\end{figure}

As before, we then explore the effect of dispersing polymers in the surrounding fluid (Fig.~\ref{fig:evolution}) and find that the polymers cause a delay in the formation of the instabilities, and, for large values of viscosity $\nu_C$, these deformations are completely suppressed.
To quantify this, we calculate the perimeter of the active region $P_C$ as a simple measure of the deformations and compare it to that of the no-polymer case $P_0$ at different points in time (Fig.~\ref{fig:peri}).
Values of $P_c/P_0<1$ indicate that the polymer case is less deformed than the no-polymer case.
Stripes display more deformation for larger values of the polymer relaxation time $\tau_C$  and smaller values of polymer viscosity $\nu_C$.
Except for the cases $\nu_C=5,10$ where there are no observed deformations, the perimeter in the polymer cases eventually approaches the value $P_0(t)$ when the stripes start to stabilize after roughly $3\times10^4$ LB time steps.

\begin{figure}[t]
\centering
\includegraphics[width=0.46\textwidth]{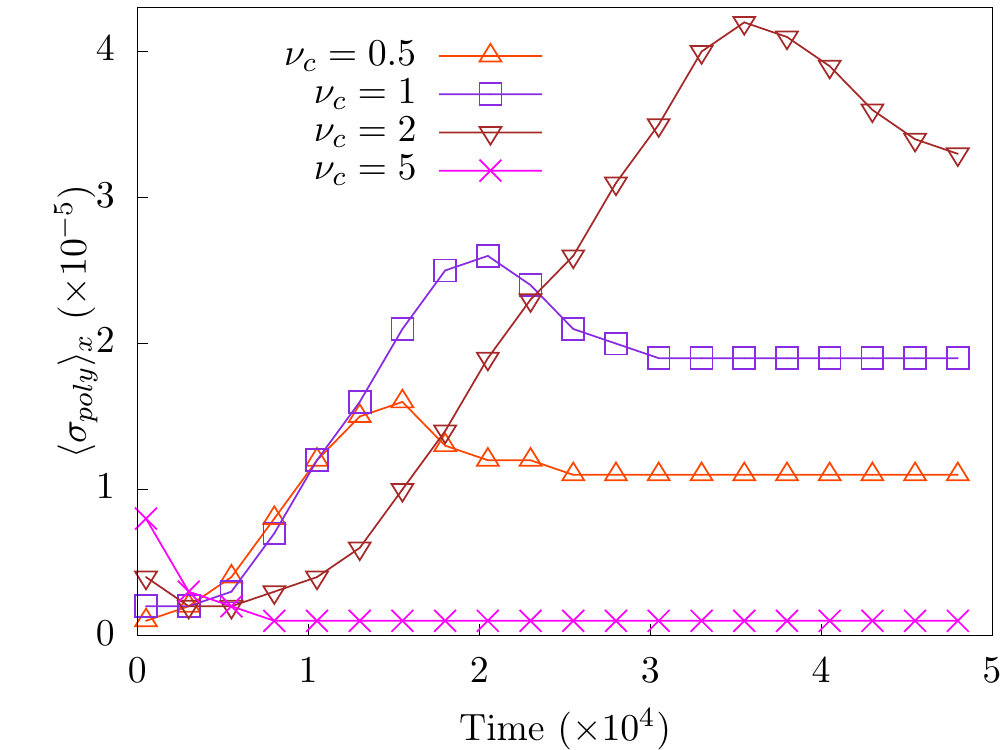}
\caption{Time series of the average polymer stress for an active stripe within a viscoelastic fluid for different values of polymer viscosity $\nu_C$.}
\label{fig:pstress_stripe}
\end{figure}

These results are consistent with how polymers decrease the magnitudes of the velocity field, in particular their hampering effect as a function of either $\tau_C$ and $\nu_C$.
It is however important to explain how no deformation of the active stripe is observed in highly viscous polymeric fluids.
Indeed, how can a highly-viscous polymeric fluid that is initially at equilibrium ($\bmC=\bmI$) suppress activity and prevent deformations, when polymer stress $\bm\sigma_{\rm polymer}$ is non-zero only when polymers are stretched ($\bmC\neq\bmI$)?
To answer this question we measure the time series of $\bm\sigma_{\rm polymer}$ in Fig.~\ref{fig:pstress_stripe}.
It is evident that velocity gradients in fact appear and stretch polymers but only at initial times.
Since polymer stresses are larger in more viscous environments (pink curve with $\nu_C=5$), they strongly damp the initial velocities and the dynamics quickly die out.
The highly-elastic fluid thus strongly damps the initial velocities and kills subsequent dynamics.
By contrast, $\bm\sigma_{\rm polymer}$ for lower values of $\nu_C$ simply reduces the velocities but does not prevent  bend instabilities from eventually occuring.

\section{Conclusion}
In this work we present a two-phase model for studying the interaction of active matter with a surrounding viscoelastic medium.
The active phase represents subcellular filaments inside the cell that are continuously put in motion by motor proteins and that exert active stresses on a surrounding viscoelastic phase, which is modeled by a polymeric fluid and is characterised by the polymer contribution to viscosity and the polymer relaxation time.
Our formulation distinguishes between the two phases in contrast to an active viscoelastic fluid \cite{HMBMRFC15,HCF16} that exhibits activity and local straining through the whole domain. 
Such a distinction is instrumental in limiting polymer elongation to near to the active-viscoelastic interface, allowing us to capture the damping effect of a polymeric fluid on active matter dynamics.
Specifically we apply the model to capture cellular dynamics, such as cell division in stiff hydrogels and cell motility of a single keratocyte cell in a mucous fluid, and the formation of instabilities in generic active matter systems.
In all cases the prominent role of environmental viscoelasticity is to suppress the flows at the interface. 
Damping is enhanced by increasing the polymer viscosity to the extent of preventing cell division, hampering cell motility, and completely suppressing hydrodynamic instabilities. 
We also report the non-trivial impact of polymer relaxation on the dynamics of active matter in a low-Deborah-number regime \Em{at fixed polymer viscosity}: the introduction of quickly-relaxing polymers in the fluid slows down the dynamics, however an increase in the polymer relaxation time leads to a behaviour that resembles active matter dynamics in an isotropic fluid medium.

Several improvements can be envisaged for further development of the current two-phase model. 
\Em{Natural polymer gels can exhibit complex stress relaxation} \cite{SPMLJ05,HRXMKLBG18} and may require polymer models more sophisticated than the Oldroyd-B model.
Moreover, many living systems interact with the viscoelastic surrounding in three dimensions. The current model can be trivially extended to three-dimensional setups, although methods to ensure efficient computation and numerical stability must be in place.
Specifically for the study of cell division, the contracting effect of the actymyosin ring can be modeled more accurately by enforcing an additional shape-dependent elongational stress around the centre of mass \cite{MYD19} instead of simply amplifying local active stress.
Lastly, an interesting research direction would be to extend this model to apply to multi-cellular organisms to take into account motility, elasticity and morphology, which are of practical importance in cancer research.
In particular it would be useful to have a more practical model that incorporates space- or time-dependent relaxation times or multiple phase fields that model other microscopic cells or tissues around a cancerous cell.

This work received funding from the Horizon 2020 research and innovation programme of the EU under Grant Agreement No. 665440. A.D. was supported by the Novo Nordisk Foundation (grant agreement No. NNF18SA0035142).

\end{document}